# Orbital selectivity of layer resolved tunneling on iron superconductor $Ba_{0.6}K_{0.4}Fe_2As_2$

**Authors:** J.-X. Yin[1]*, X. -X. Wu[1]*, Jian Li[2], Zheng Wu[2], J.-H. Wang[2], C.-S. Ting[2], P.-H. Hor[2], X. J. Liang[1], C. L. Zhang[3], P. C. Dai[3], X. C. Wang[1], C. Q. Jin[1,8], G. F. Chen[1], J. P. Hu[1,4], Z. -Q. Wang[5], Ang Li[2,6,7], H. Ding[1,8,9], S. H. Pan[1,2,8,9,10]†

**Affiliations:**

[1]Institute of Physics, Chinese Academy of Sciences, Beijing 100190, China.

[2]TCSUH and Department of Physics, University of Houston, Houston, Texas 77204, USA.

[3]Department of Physics and Astronomy, Rice University, Houston, Texas 77005, USA.

[4]Department of Physics, Purdue University, West Lafayette, Indiana 47907, USA.

[5]Department of Physics, Boston College, Chestnut Hill, Massachusetts 02467, USA.

[6]State Key Laboratory of Functional Materials for Informatics, Shanghai Institute of Microsystem and Information Technology, Chinese Academy of Sciences, Shanghai 200050, China

[7]CAS Center for Excellence in Superconducting Electronics, Shanghai 200050, China

[8]School of Physics, University of Chinese Academy of Sciences, Beijing 100190, China

[9]CAS Center for Excellence in Topological Quantum Computation, University of Chinese Academy of Sciences, Beijing 100190, China

[10]Songshang Lake Material Laboratory, Dongguan, Guangdong 523808, China

*These authors contributed equally to this work.

†Corresponding author. E-mail: span@iphy.ac.cn.

**We use scanning tunneling microscopy/spectroscopy (STM/S) to elucidate the Cooper pairing of the iron pnictide superconductor $Ba_{0.6}K_{0.4}Fe_2As_2$. By a cold-cleaving technique, we obtain atomically resolved termination surfaces with different layer identities. Remarkably, we observe that the low-energy tunneling spectrum related to superconductivity has an unprecedented dependence on the layer-identity. By cross-referencing with the angle-revolved photoemission results and the tunneling data of LiFeAs, we find that tunneling on each termination surface probes superconductivity through selecting distinct Fe-3*d* orbitals. These findings imply the real-space orbital features of the Cooper pairing in the iron pnictide superconductors, and propose a new and general concept that, for complex multi-orbital material, tunneling on different terminating layers can feature orbital selectivity.**

Iron pnictide superconductors share a common structure based on a planar layer of Fe atoms tetrahedrally coordinated by As anions. Analogous to the key role of the copper-oxide plane plays in cuprates superconductivity, the iron-arsenic trilayer is the essential structure responsible for the emergence of superconductivity in the iron pnictides. Unlike the copper-oxide plane, which has a single Fermi sheet involving the Cu $d_{x^2-y^2}$ orbital, the iron-arsenic trilayer has multiple Fermi sheets involving all of the $t_{2g}$ Fe *d*-orbitals [1-7]. Consequently, the orbital nature of the electron pairing becomes one of the essential issues in understanding iron-based superconductivity. K-doped $BaFe_2As_2$ and LiFeAs are ideal materials to investigate this problem owing to their stoichiometric Fe-As structure. However, the active surface K atoms can lead to lattice disorder for K-doped Ba122 materials [8-10], while native defects are easily introduced during the growth process of the LiFeAs materials [11-13].

Both disorder and native defects can affect the Fe-As structural integrity and alter the intrinsic superconducting properties at the surface [14]. Using cryogenic *in situ* cleaving technique, we have been able to obtain ordered surfaces with definite atomic identities in $Ba_{0.6}K_{0.4}Fe_2As_2$ [15]. By finely controlling the growth rate, we can synthesize LiFeAs single crystals with large, defect-free areas. A high-resolution STS measurement of the superconducting gap on these ordered and clean surfaces is of great importance to understand the intrinsic electron pairing.

As illustrated in Fig. 1 (a) and (b), cleavage of $Ba_{0.6}K_{0.4}Fe_2As_2$ single crystals ($T_C$=38K) mainly breaks the As-Ba(K) bonding, generating the $\sqrt{2}\times\sqrt{2}R45°$ bucking reconstructed Ba(K)-terminating layer (left upper corner in Fig. 1(b)) or the dimerized As-terminating layer (left lower corner in Fig. 1(b)) [15]. Decisive experimental evidence for such surface identification was found by imaging the symmetry-dictated surface boundary and the layer-selective chemical dopants in Ref. 15. In addition to these two ordered surfaces, we also observe a disordered surface (right lower corner in Fig. 1b). This disorder is likely caused by the active K atoms, as a mixture of Ba atoms and K atoms, because no such surface is observed in the $BaFe_2As_2$ samples in our systematic study. The disordered surface exhibits an inhomogeneous superconducting gap that varies from 2 meV to 10 meV in the gap-map and line-cut spectrums (Figs. 1c-e), which is similar to previous STM/S results [8-10]. Here we note that the superconducting gap size is measured from the distance between coherent peaks in the tunneling spectrum. However, such severe inhomogeneity is inconsistent with the results on other iron-based superconductors with ordered terminating surfaces [14], and the obtained gap sizes are much smaller than those obtained from our angle-resolved photoemission spectroscopy (ARPES) measurements on the same material [16,17], while smaller gap size was reported in a laser based ARPES experiment [18].

In contrast to the disordered surface, the gap magnitudes observed on the As-terminating and Ba(K)-terminating layers are fairly homogenous in both cases. More remarkably, their gap structures are strikingly different from each other. Along the line drawn in Fig. 2(a), we take spectrums from the Ba(K)-terminating region to the As-terminating region. As can be seen in Fig. 2(b), the superconducting gap is 6.0 meV deep inside the Ba(K)-terminating region and 10.5 meV deep within the As-terminating region. From the gap-maps taken on the As and the Ba(K) regions (Figs. 2(c) and (d)), we also find that the gap size varies less than 1.0 meV within each region, demonstrating the intrinsic spatial homogeneity of the electron pairing strength. To explore the fine structures of the energy gap on the ordered surfaces, we measure the spectra at a lower temperature, as displayed in Fig. 2(e). Both spectrums have flat bottoms indicating the full energy gaps with no density of states near the Fermi energy, which directly excludes any nodal symmetry of the electron pairing. Taking a closer look at the spectra on the As surfaces, we can find that there are extra shoulders around ±6 meV as marked by the blue arrows. For the spectra on the Ba(K) surfaces, we can also find that there are very weak peaks s around ±10.5 meV indicated by green arrows. These subtle features, together with the strong coherent peaks illustrate a two-gap structure, but with different spectral weight distributions on their respective terminating layers. We also note that there are additional bumps at ±18meV in the tunneling spectra taken on As and Ba(K) surfaces, whose origin is currently unclear.

Compared with $BaFe_2As_2$, cleaved LiFeAs ($T_C$=17.5K) has only one termination surface, i.e. the Li layer (Fig.3(a)). In Fig. 3(b) we present an image of a large defect-free Li surface. With high-resolution zoomed-in imaging, both Li and As atoms can be captured [19]. We note that the observation of both atoms is consistent with the subatomic distance between the As and Li layers (~0.5 Å in bulk). A two-gap structure is clearly displayed on this surface. The line-cut spectra shown in Fig. 3(c) demonstrate an extremely homogenous and fully-opened gap in this system [11][12][13][20][21]. Measuring at a much lower temperature, we observe a clear two-gap structure as shown in Fig. 3(d).

To directly associate these gap structures with the iron-based superconductivity, we re-plot the dI/dV spectrums on each clean surface in the energy unit of $k_B T_C$ in Fig. 4(a). For both materials, the ratio $2\Delta/k_B T_C$ of the small gap is around 4 and of the large gap is around 7, suggesting a strong coupling superconducting ground state [22]. The fact of the similar ratios of $2\Delta/k_B T_C$ in these two different materials supports that the tunneling spectrums on the ordered and atomically clean surfaces largely reflect the intrinsic electron pairing within the Fe-As layer. We then compare our results with the gap structures measured in momentum space. Considering that the c-axis tunneling states mostly carry a small in-plane momentum, we focus only on the two hole-like bands near the center of the Brillouin zone for comparison. Figure. 4(b) shows the symmetrized spectral intensity of these two bands [16,17] measured by our ARPES, also marked with their dominant orbital characters [5-7]. Although both Ba(K) and As terminating surfaces have reconstructions, the gap values given by the associated tunneling spectra are well consistent with that of the $k_Z$-averaged ARPES data, which measures the bulk property. Thus, the surface reconstructions do not seem to substantially alter the Cooper pairing, and the termination dependent tunneling signal can result from their orbital selectivity based on the orbital distinctions in the ARPES data. The direct match of the superconducting gap size suggests that the tunneling on the Ba(K), As, and Li-As surface selects the $d_{xy}$, $d_{xz/yz}$, and all the $t_{2g}$ orbitals, respectively.

In the following, we discuss the interpretation of the observed orbital-selectivity. Based on the symmetry group of the bulk Fe-As trilayer [23] as $Z_2 \otimes D_{2d}$, Fe $3d_{xz/yz}$ mainly hybridizes with As $4p_{x/y}$ orbital while Fe $3d_{xy}$ with As $4p_z$ orbital [23,24]. Accordingly, for the Ba(K) terminating layer, which is located significantly above the As plane, the tunneling states come mainly from the Ba(K) extended 6*s* orbital overlapping with the As $p_z$ orbital, thus the tunneling electrons are mostly from Fe $3d_{xy}$ orbital. In contrast, the As terminating layer has a dimer row reconstruction, where the $4p_z$ orbitals of the As dimer form a π bonding state and a π* antibonding state. We find, from first-principles calculations, that the surface As $4p_z$ orbital hybridizes with the Fe $3d_{xz/yz}$ orbital, which is related to the π* antibonding state formed near $E_F$ (Supplementary Material [25]). Therefore, the tunneling electrons are mainly from the Fe $3d_{xz/yz}$ orbital. For the Li-As terminating surface in LiFeAs, due to the much smaller interlayer distance between Li and As layers, tunneling can occur through both Li and As states, thus selecting all Fe $t_{2g}$ orbitals.

The above discussion sketches a consistent picture for interpreting our experimental observations on $Ba_{0.6}K_{0.4}Fe_2As_2$ and LiFeAs. Crucially, this simple physical picture of the orbital selectivity appears to also be valid in the understanding of other STM/S results on iron pnictides. For example, a shallower and less coherent superconducting gap spectrum is observed on Ba surface compared with that on As surface in the Co-doped Ba122 system [15]. Based on the orbital selectivity picture, the tunneling on Ba surface mainly probes the $d_{xy}$ orbital, then the less coherent gap spectrum indicates that the decoherence effect of the scattering due to Co dopants is mainly in the $d_{xy}$ orbital channel, which is consistent with the ARPES results [6,7]. There are many other examples, two of such are the quasi-particle interference of the As dimer-row surface in the Co-doped Ca122 system associated with the intraband scattering of the inner hole pocket ($d_{xz/yz}$ orbital) [31] and the tunneling spectrum on the √2×√2R45° reconstructed surface of $KFe_2As_2$ exhibiting a Van Hove singularity, which features $d_{xy}$ orbital in ARPES [22]. It is expected that a more comprehensive theoretical analysis considering both the tunneling matrix element and the Wannier-function-based first-principles simulations shall greatly substantiate the general applicability of this orbital selective picture.

In conclusion, we have performed layer resolved STM/STS experiments on iron pnictide superconductors $Ba_{0.6}K_{0.4}Fe_2As_2$ single crystals and have found that the tunneling spectra are highly terminating-surface dependent. Discussions of our experimental observations point to the real-space orbital features of the Cooper pairing in the iron pnictide superconductors. Moreover, this research work, correlated with the ARPES results, unifies the multi-orbital nature of iron-based superconductivity in the real and momentum space, which can be relevant to some advanced concepts including orbital selective Cooper pairing [33][34], orbital selective correlation [35] and Hund's superconductivity [36]. It would be interesting to see how the tunneling signals on different surfaces response to the external magnetic field [37][38][39] in future experiment. Finally, this research work also proposes the general concept that layer resolved tunneling can feature orbital selectivity, which will be very useful in the study of complex layered materials [39][40][41][42][43][44][45][46].

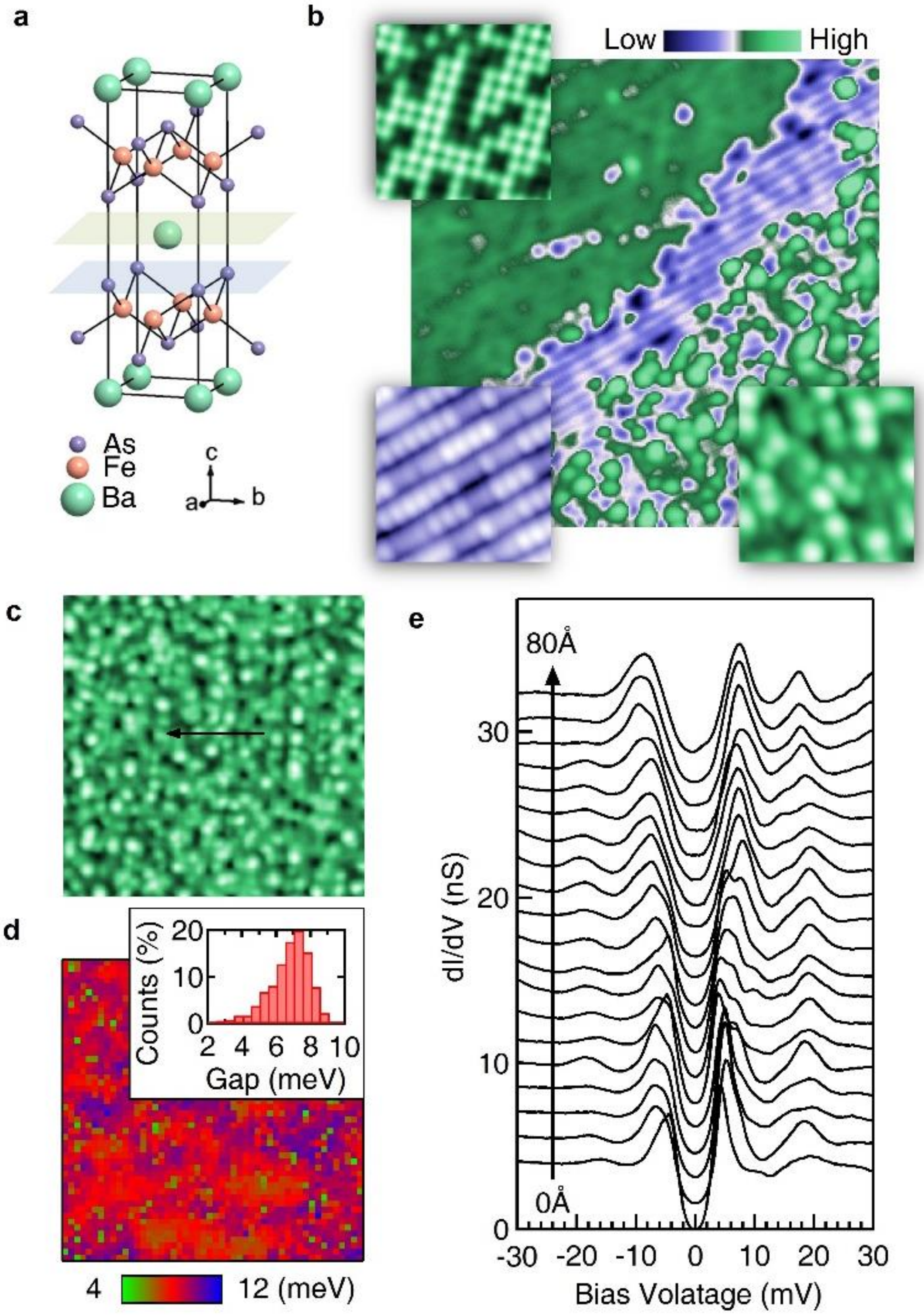

FIG. 1 (a) Crystal structure of $BaFe_2As_2$. (b) A topographic image shows three kinds of surfaces (V = -100 mV, I = 0.3 nA, 300 × 300 Å). A zoom-in image at the left upper corner shows the Ba(K) surface (V = -50 mV, I = 0.3 nA, 50×50 Å). A zoom-in image at the left lower corner shows the As surface (V = -30 mV, I = 0.3 nA, 50 × 50 Å). A zoom-in image at the right lower corner shows the disordered surface (V = -50 mV, I = 0.3 nA, 50 × 50 Å). (c) (d) Topographic image of a disordered surface (V = -100 mV, I = 0.03 nA, 300 × 300 Å) and its gap-map (V = -20 mV, I = 1nA), respectively. The superconducting gap at each point is determined by the distance between coherent peaks. (e) Line-cut spectra along the arrow-line in (c). Spectra are offset for clarity. All data are acquired at 4.2 K.

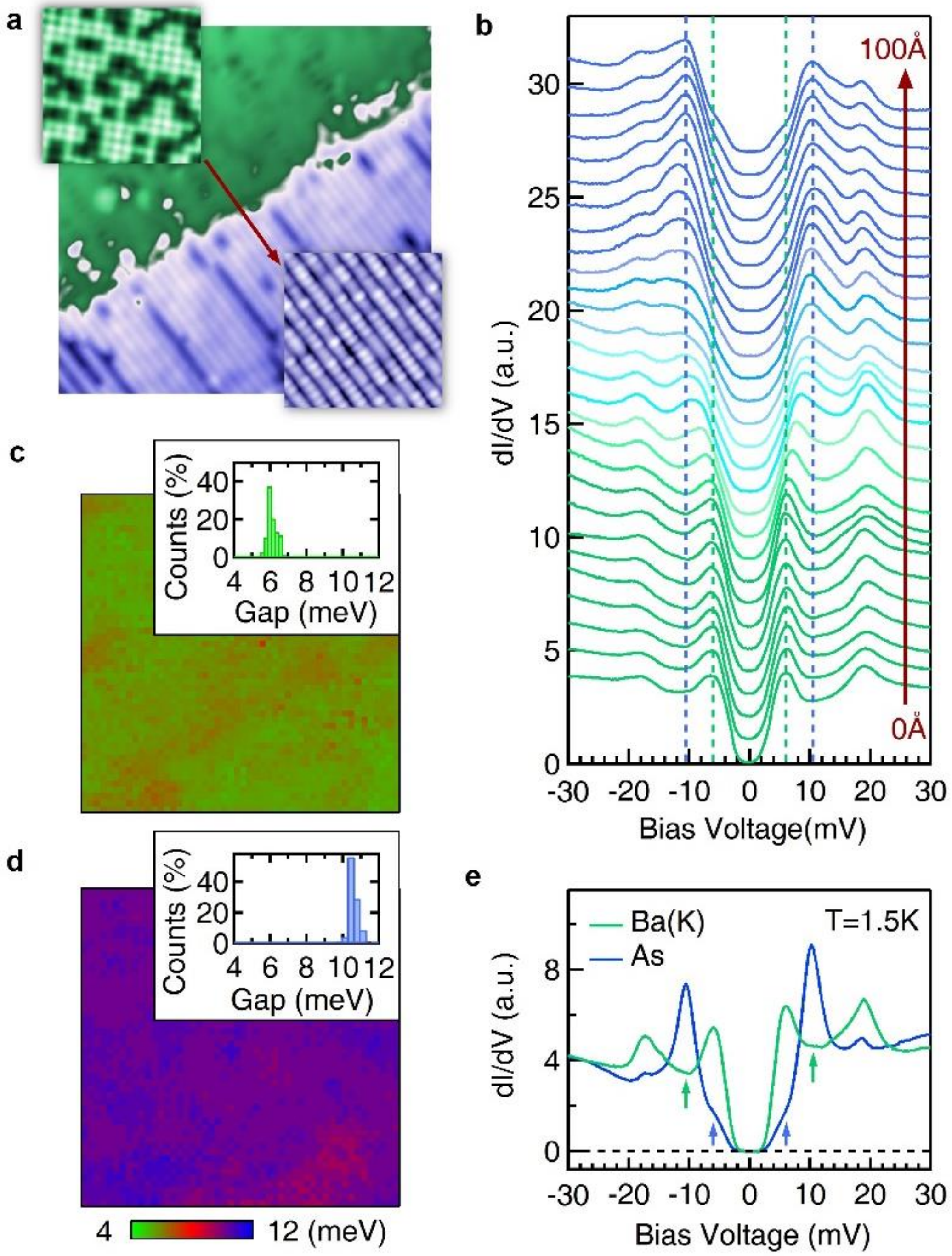

FIG. 2 (a) Topographic image near the boundary between the ordered Ba(K) surface (left upper area) and As surface (right lower area) (V = -100 mV, I = 0.03 nA, 300 × 300 Å, T = 4.2K). A zoom-in image at the left upper corner shows the Ba(K) surface (V = -50 mV, I = 0.3 nA, 80 × 80 Å). A zoom-in image at the right lower corner shows the As dimmer row surface (V = -30 mV, I = 0.3 nA, 80 × 80 Å). (b) Line-cut spectra through the surface boundary as marked in (a). Spectra are offset for clarity. (c) (d) Gap-maps for the ordered Ba(K) and As surface (V = -30 mV, I = 1 nA, 100 × 100 Å, T = 4.2K), respectively. **e,** Spectrums taken on the Ba(K) and As surfaces, respectively (V = -30 mV, I = 1 nA, T = 1.5 K).

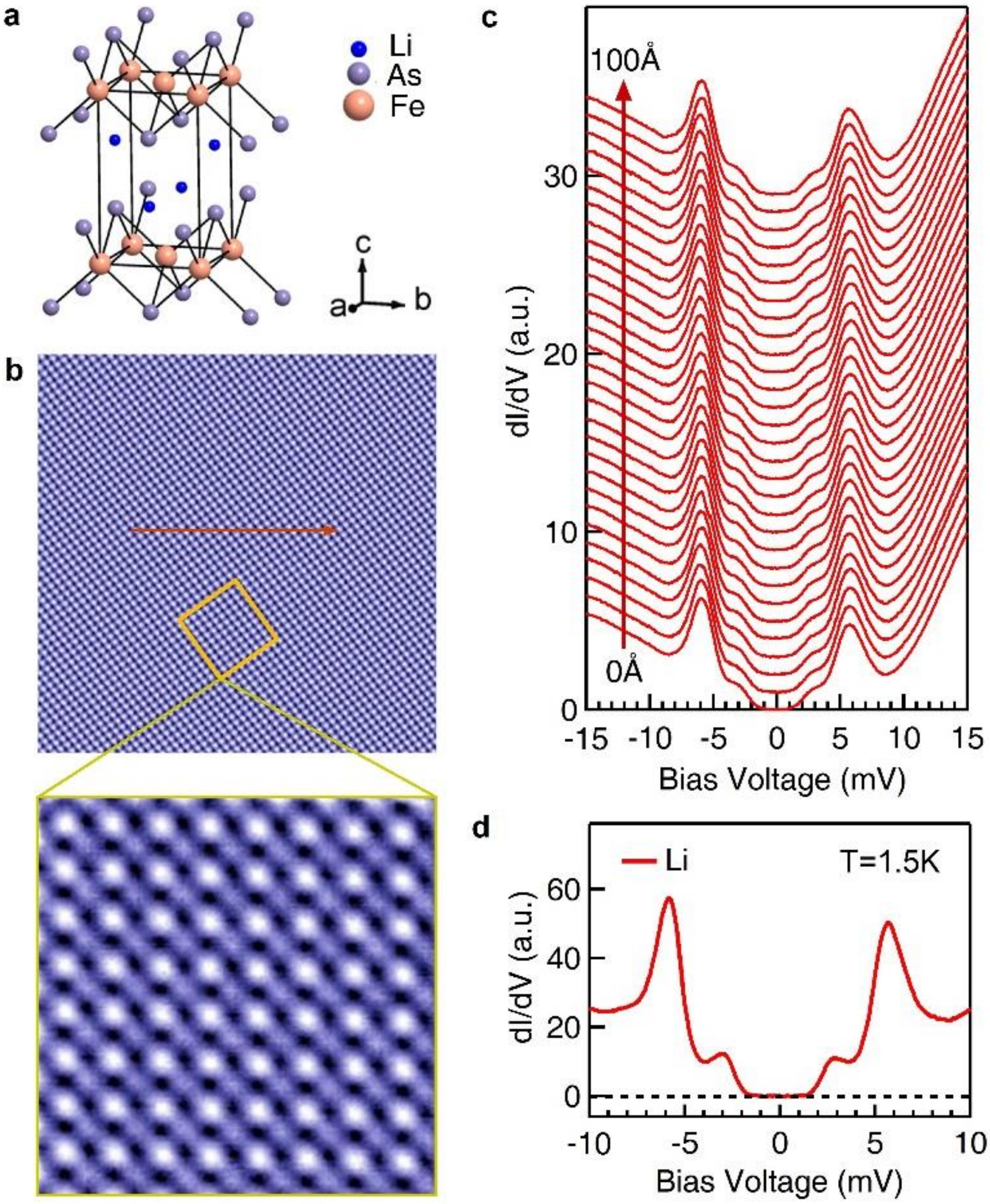

FIG. 3 (a) Crystal structure for LiFeAs. (b) Topographic image of the Li surface (V = -100 mV, I = 1 nA, 200 × 200 Å, T = 4.2 K). High resolution zoom-in image showing both the Li and As atoms (V = -10 mV, I = 0.5 nA, 40 × 40 Å, T = 4.2 K). (c) Line-cut spectra taken on the Li surface in (b) (V = -15 mV, I = 0.5 nA, T = 4.2 K). Spectra are offset for clarity. (d) Spectrum taken on the Li surface showing a double-gap feature (V = -15 mV, I = 0.5 nA, T = 1.5 K).

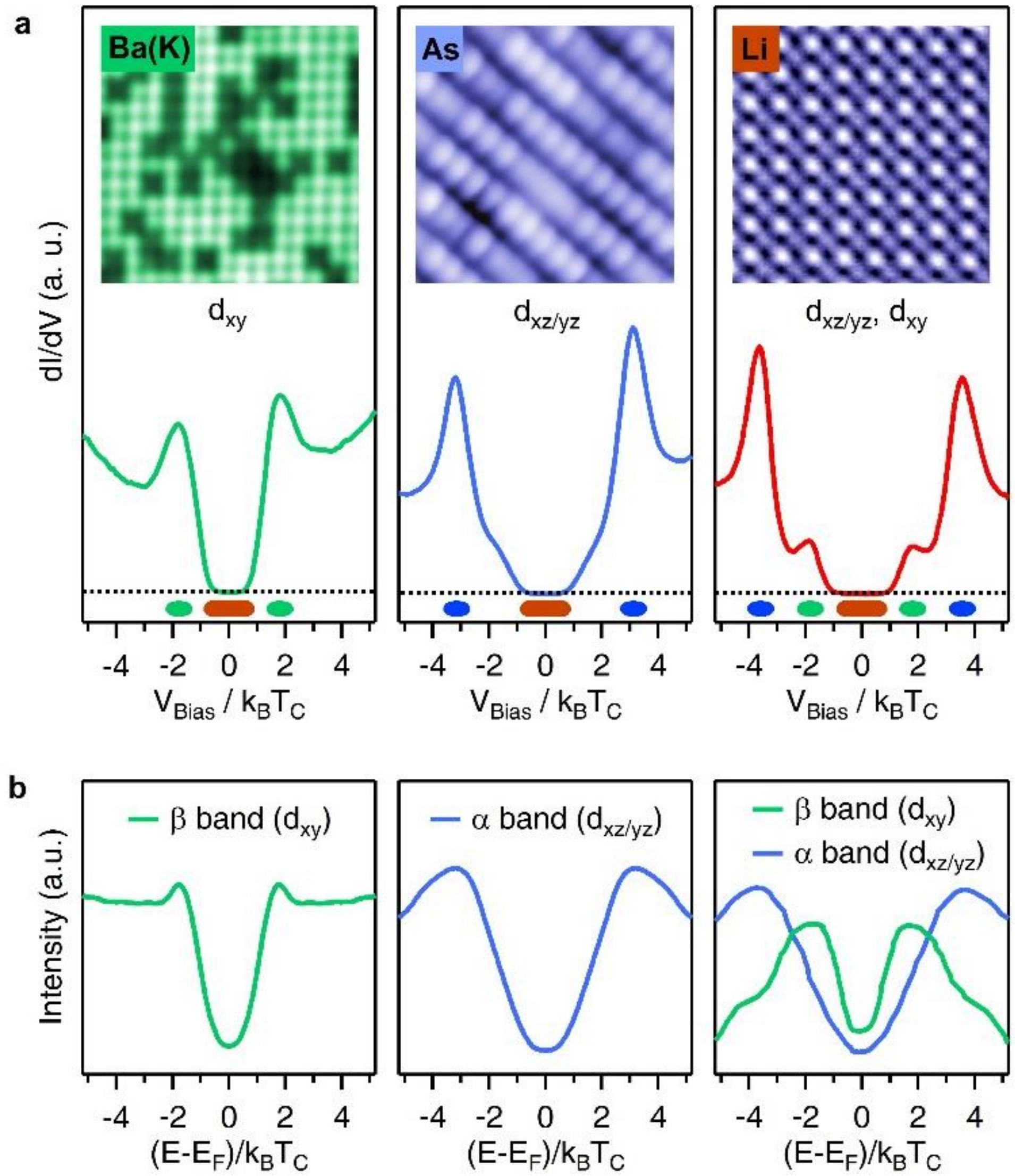

FIG. 4 (a) STS spectra in the energy unit of $k_BT_C$ on three different kinds of surfaces taken at 1.5 K. The red bars mark the flat bottom range, the green/blue bars estimate the anisotropy range of the small/large gap. The inset images show the corresponding surfaces. (b) Symmetrized ARPES spectra in the energy unit of $k_BT_C$ for β band (outer hole pockets) in $Ba_{0.6}K_{0.4}Fe_2As_2$ (T = 10 K [17]), α band (inner hole pockets) in $Ba_{0.6}K_{0.4}Fe_2As_2$ ($k_Z$ averaged, T = 10 K [16]), and both bands in LiFeAs (T = 8 K [17]), respectively.

**Acknowledgments:**

The authors thank Tao Xiang, D.-H. Lee, T.-K. Lee, Z.Y. Lu, Sen Zhou, S.-F. Wu and Hu Miao for stimulating discussions. This work is supported by the Chinese Academy of Sciences, NSFC (11227903), BM-STC (Z181100004218007), the Ministry of Science and Technology of China (2015CB921300, 2015CB921304,2017YFA0302903), the Strategic Priority Research ProgramB (XDB04040300, XDB07000000). A.L. acknowledges support from the Shanghai Science and Technology Committee(18ZR1447300). Work at University of Houston was sup-ported by the State of Texas through TcSUH. C.-S.T. acknowl-edges support of the Robert A. Welch Foundation (E-1146).Work at ORNL was supported by the U. S. Department ofEnergy, Office of Science, Basic Energy Sciences, MaterialsSciences and Engineering Division.